\documentclass[10pt,english]{article}
\usepackage{subfigure}
\usepackage{graphicx}
\usepackage{float}
\usepackage[T1]{fontenc}
\usepackage[latin9]{inputenc}
\usepackage[margin=1.4in]{geometry}
\usepackage{float}
\usepackage{amsmath}
\usepackage{amsbsy}

\usepackage{setspace}
\usepackage{amssymb}
\usepackage{esint}
\usepackage{cite}
\newfloat{algorithm}{tbp}{loa}
\floatname{algorithm}{Algorithm}
\usepackage{algorithmic}

\newlength\myindent
\newcommand\bindent{
  \begingroup
  \setlength{\itemindent}{\myindent}
  \addtolength{\algorithmicindent}{\myindent}
}
\newcommand\eindent{\endgroup}

\makeatletter

\floatstyle{ruled}
\newfloat{algorithm}{tbp}{loa}
\floatname{algorithm}{Algorithm}

\usepackage{babel}

\begin{document}

\title{Fast Time Series Detrending with Applications to\\ Heart Rate Variability Analysis}

\author{M. Andrecut}

\date{June 21, 2019}

\maketitle
{

\centering Calgary, Alberta, T3G 5Y8, Canada

\centering mircea.andrecut@gmail.com

} 

\begin{abstract}
Here we discuss a new fast detrending method for the non-stationary RR time series used in Heart Rate Variability analysis. 
The described method is based on the diffusion equation, and we show numerically that it is equivalent to the widely used Smoothing Priors Approach (SPA)
and Wavelet Smoothing Approach (WSA) methods. The speed of the proposed method is comparable to the WSA method and it is several orders of magnitude faster than the SPA method, 
which makes it suitable for very long time series analysis. 



\end{abstract}

\section{Introduction}

Heart Rate Variability (HRV) is an important indicator used in the analysis of the autonomous nervous system controlling 
the cardiovascular activity \cite{key-1}. The HRV is extracted from the fluctuations of the heart beat intervals (RR intervals) obtained 
from electrocardiogram (ECG) measurements, and it is used for the assessment of different patient groups at risk of 
cardiac diseases, such as: lifethreatening arrhythmias, myocardial infarction, and diabetes mellitus \cite{key-1,key-2}.

The HRV analysis requires reliable and efficient signal processing techniques of the series of RR-intervals. 
Both time domain and frequency domain analysis can be used to extract useful clinical information 
from the ECG measurements and the corresponding RR time series \cite{key-3,key-4}.  However, the extracted RR time series are non-stationary, 
and contain trends and artifacts that must be removed before analysis. 
Especially, the frequency domain analysis is sensitive to the trends embedded in the RR 
time series \cite{key-3,key-4}. These trends have a complex behavior and are caused by external effects, corresponding to 
smooth and slowly varying processes, which can have both oscillating and stochastic components.
Several methods have been developed to remove these complex trends from the RR time series. Most methods are based on 
polynomial regression models \cite{key-5}, spectral filters (wavelets etc.) \cite{key-6}, and empirical mode decomposition \cite{key-7}. 

One of the most reliable and widely used detrending method is based on the Smoothing Prior Approach (SPA), 
which acts as a time-varying finite impulse high-pass filter in the frequency domain \cite{key-8}. 
Another frequently used method is based on the Wavelet Smoothing Approach (WSA) \cite{key-9}, which is probably the fastest method currently in use.  

In this paper we discuss a new fast detrending method based on the diffusion equation, and we show numerically that it is 
equivalent to the SPA and WSA methods. However, our numerical implementation shows that the speed of the diffusion equation approach is comparable 
to the WSA method, and therefore it is several orders of magnitude faster than the SPA method, which makes it suitable for very long time series analysis. 
While here we limit our discussion to the HRV analysis, we should mention that the described method can be used to accelerate any 
time series detrending, smoothing or fitting tasks.

\section{SPA detrending}

Let us denote the RR time series by:
\begin{equation}
r = [R_1-R_0,R_2-R_1,...,R_N-R_{N-1}]^T \in \mathbb{R}^N,
\end{equation}
where $N$ is the number of $R$ peaks detected. The RR time series is considered to have two components \cite{key-8}:
\begin{equation}
r = x + y,
\end{equation}
where $x$ is the nearly stationary component, and $y$ is the slowly varying trend. 
Typically the trend component can be approximated with a linear model such as:
\begin{equation}
y = H \theta + \eta,
\end{equation}
where $H$ is the observation matrix, $\theta$ are the regression parameters and $\eta$ is the observation error. 
The corresponding regularized least squares problem is:
\begin{equation}
\theta(\mu) = \text{arg} \min_{\theta} \lbrace \Vert H\theta -r \Vert^2 + \mu \Vert D_d (H \theta) \Vert^2 \rbrace,
\end{equation}
where $\mu>0$ is the regularization parameter and $D_d$ is the discrete approximation of the $d$-th derivative operator.
The solution of this problem is given by:
\begin{equation}
\theta(\mu) = (H^T H + \mu H^T D_d^T D_d H)^{-1} H^T r,
\end{equation}
such that:
\begin{equation}
y(\mu) = H \theta(\mu).
\end{equation}
The observation matrix $H$ can be calculated using different regression functions like: polynomials, Gaussians or sigmoids. 
However, in order to avoid the selection problem of the basis functions, the SPA simply uses $H=I$, where $I\in \mathbb{R}^{N\times N}$ 
is the identity function \cite{key-8}.  Also, SPA uses the second order derivative operator $D_2 \in \mathbb{R}^{(N-2)\times N}$ 
in order to estimate the regularization part of the optimization problem:\cite{key-8}
\begin{equation}
D_2 = \begin{bmatrix}
    1 & -2 & 1 & 0 & \dots  & 0 \\
    0 & 1 & -2 & 1 & \dots  & 0 \\
    \vdots & \ddots & \ddots & \ddots & \ddots & \vdots \\
    0 & \dots 0 & 0 & 1 & -2  & 1
\end{bmatrix}.
\end{equation}
With these SPA parameters, the detrended RR time series is given by:
\begin{equation}
x(\mu) = r - y(\mu) = [ I - (I + \mu D_2^T D_2)^{-1}] r.
\end{equation}

A direct implementation of the SPA method using the above equation is quite costly because the matrix inversion, which is an $O(N^3)$ operation. 
A better approach is to speculate the fact that the matrix is symmetrical, and therefore one can use the Cholesky decomposition:
\begin{equation}
I + \mu D_2^T D_2 = LL^T,
\end{equation}
where $L$ is a lower triangular matrix. Then the problem reduces to successively solving two triangular systems of equations:
\begin{equation}
Lr' = r \Rightarrow L^Ty=r',
\end{equation}
in order to extract the trend $y$.

\section{WSA detrending}

WSA detrending is a spectral method using the discrete wavelet decomposition of the RR time series \cite{key-9,key-10}. 
Wavelets provide an analysis of the signal which is localized in both time and frequency, unlike the Fourier transform which is localized only in frequency. 
The continuous wavelet transform (CWT) of $r(t)$ is defined as the convolution:
\begin{equation}
\hat{r}(a,b) = \frac{1}{\sqrt{a}}\int_{-\infty}^{\infty}\Psi\left( \frac{t-b}{a} \right) dt,
\end{equation}
where $\Psi(t)$ is the mother wavelet, depending on two parameters: the scale dilation $a$ and the translation $b$. 
We should note that there are an infinite number of choices of $\Psi(t)$ functions. 

In computational tasks one can use the discrete wavelet decomposition (DWT), which is a sampled version of CWT, 
transforming a discrete signal into discrete coefficients in the wavelet domain. Thus, instead of using 
the continuous coefficient values $a,b\in \mathbb{R}$, the values $\hat{r}(a,b)$ are calculated over a dyadic discrete grid:
\begin{equation}
\hat{r}_{j,k}= \frac{1}{\sqrt{2^j}}\int_{-\infty}^{\infty}\Psi\left( \frac{t-k2^j}{2^j} \right) dt, \quad j,k \in \mathbb{Z}.
\end{equation}

The DWT can be computed efficiently using the Mallat's algorithm, which has a complexity $O(N)$ \cite{key-11}. The algorithm decomposes the 
signal into different scales, corresponding to different frequency bands, using low-pass (LP) and respectively high-pass (HP) filters. 
The output values of the LP filters correspond to the approximation coefficients, while the output values of the HP filters 
correspond to the detail coefficients. The number of decomposition levels (typically 3 to 5) is a parameter depending on the signal to be analyzed. 

Due to the orthogonality of the wavelets, the DWT has the interesting property of transforming noise into noise \cite{key-9}. 
This property can be used to smooth the signal, and to extract the trend by keeping only the coefficients which are stronger 
than the noise level in the signal. This procedure is called wavelet thresholding and consists in comparing the coefficients 
with a threshold in order to decide if they are included in the inverse reconstruction process. 
Thus, wavelet thresholding is setting to zero the coefficients with an absolute value below a certain threshold level $\tau$, 
using the hard (H) or the soft (S) thresholding functions:
\begin{equation}
\Theta^H_{\tau}(x) = 
  \begin{cases}
   0  & \text{if } |x| \leq \tau \\
   x & \text{if } |x| > \tau
  \end{cases}, \quad
\Theta^S_{\tau}(x) = 
  \begin{cases}
   0  & \text{if } |x| \leq \tau \\
   x - \tau & \text{if } x > \tau \\
   x + \tau & \text{if } x < -\tau
  \end{cases}. 
\end{equation}

A frequently used estimate of the optimal threshold $\tau$ value is the universal threshold which can be used with both hard and soft thresholding functions \cite{key-12}: 
\begin{equation}
\tau = \sigma \sqrt{2\ln N},
\end{equation}
where $\sigma$ is the standard deviation of the detail wavelet coefficients. 
We should note that the thresholding procedure is usually applied only to the detail coefficients \cite{key-13}.
After applying the thresholding procedure, the trend can be recovered using the inverse DWT. 

\section{Diffusion detrending}

Let us rewrite the RR time series as:
\begin{equation}
r = [r_0,r_1,...,r_{N-1}]^T \in \mathbb{R}^N, 
\end{equation}
where $r_i=R_{i+1}-R_i$. Also, we consider the following optimization problem:
\begin{equation}
y(\mu) = \text{arg} \min_{y} S(y,\mu),
\end{equation}
where:
\begin{equation}
S(y,\mu) = F(y) + \mu G(y), 
\end{equation}
with:
\begin{equation}
F(y) = \sum_i (y_i - r_i)^2,
\end{equation}
and respectively:
\begin{equation}
G(y) = \sum_i (y_{i-1} - 2y_i + y_{i+1})^2.
\end{equation}
Obviously this problem is equivalent to the SPA optimization problem.

Let us now consider the following diffusion equation:
\begin{equation}
\frac{\partial y}{\partial t} = \alpha \nabla^2 y,
\end{equation}
where $\alpha$ is the diffusion coefficient. We also assume that the initial condition of the diffusion equation is:
\begin{equation}
y(0) = r \Leftrightarrow y_i(0) = r_i, i=0,1,...,N-1. 
\end{equation}
The finite difference iteration of the above equation can be written as:
\begin{equation}
y_i(t+1) = y_i(t) + \alpha [y_{i-1}(t) - 2 y_i(t) + y_{i-1}(t)] = y_i(t) + \alpha L_i(t),
\end{equation}
where:
\begin{equation}
\frac{\partial y}{\partial t} \biggr\rvert_i \simeq y_i(t+1) - y_i(t),
\end{equation}
and
\begin{equation}
\partial_{y_i}^2 y \simeq y_{i-1}(t) - 2 y_i(t) + y_{i-1}(t) = L_i(t),
\end{equation}
are the discrete approximations of the the time derivative and respectively of the Laplacian. 

One can see that the gradient descent solution for the minimization of $G(y)$ is also given by:
\begin{equation}
y_i(t+1) = y_i(t) - \gamma \partial_{y_i}G(y) = y_i(t) + \gamma L_i(t), 
\end{equation}
where $\gamma > 0$. For  $\gamma=\alpha$ this is obviously equivalent to the finite difference solution of the diffusion equation, 
which therefore minimizes the objective function $G(y)$. 
Thus, starting from $y = r$, with the maximum value of the objective optimization function:
\begin{equation}
S(y(0),\mu) = \mu G(r),
\end{equation}
we let the diffusion process shrink and smooth $y$ until the minimum of $S(y,\mu)$ is reached 
for a given $\mu$. The pseudo-code of the Diffusion Detrending Algorithm (DDA) is listed in Algorithm 1. 
One can see that the maximum number of iterations is set to $N$, and the function returns the trend $y$, the 
almost stationary component $x=y-r$, and the time step $t$ when the minimum of $S$ has been reached. 
Also, the first and the last points of the trend are computed as:
\begin{align}
y_0(t+1) &= y_0(t) + 2\alpha [y_{1}(t) - y_0(t)],\\
y_{N-1}(t+1) &= y_{N-1}(t) + 2\alpha [y_{N-2}(t) - y_{N-1}(t)].
\end{align}

\begin{algorithm}[h!]
\caption{Diffusion Detrending Algorithm (DDA).}
\begin{algorithmic}
\STATE $\textbf{function} \text{ diffusion\_detrending}(r, \mu, \alpha)$
\bindent
\STATE $N, S, y, t \leftarrow \text{length}(r), \infty, r, 0$
\WHILE{$\text{true}$}
	\STATE $S^{*} \leftarrow \sum_i [(y_i - r_i)^2 + \mu(y_{i-1} - 2y_i + y_{i+1})^2]$
	\IF{$S^{*}>S$ $\textbf{or}$ $t>N$}
		\STATE $\textbf{break}$
	\ELSE
		\FOR{$i=1:N-1$}
			\STATE $y_i \leftarrow y_{i} + \alpha(y_{i-1}-2y_{i}+y_{i+1})$
		\ENDFOR
		\STATE $S,t \leftarrow S^{*},t+1$ 
		\STATE $y_0,y_{N-1} \leftarrow y_0+2\alpha(y_1-y_0), y_{N-1} + 2\alpha(y_{N-2}-y_{N-1})$ 
	\ENDIF
\ENDWHILE
\RETURN $t, y, r-y$
\eindent
\STATE $\textbf{end function}$
\end{algorithmic}
\end{algorithm}

\begin{figure}[ht!]
\centering \includegraphics[width=13.75cm]{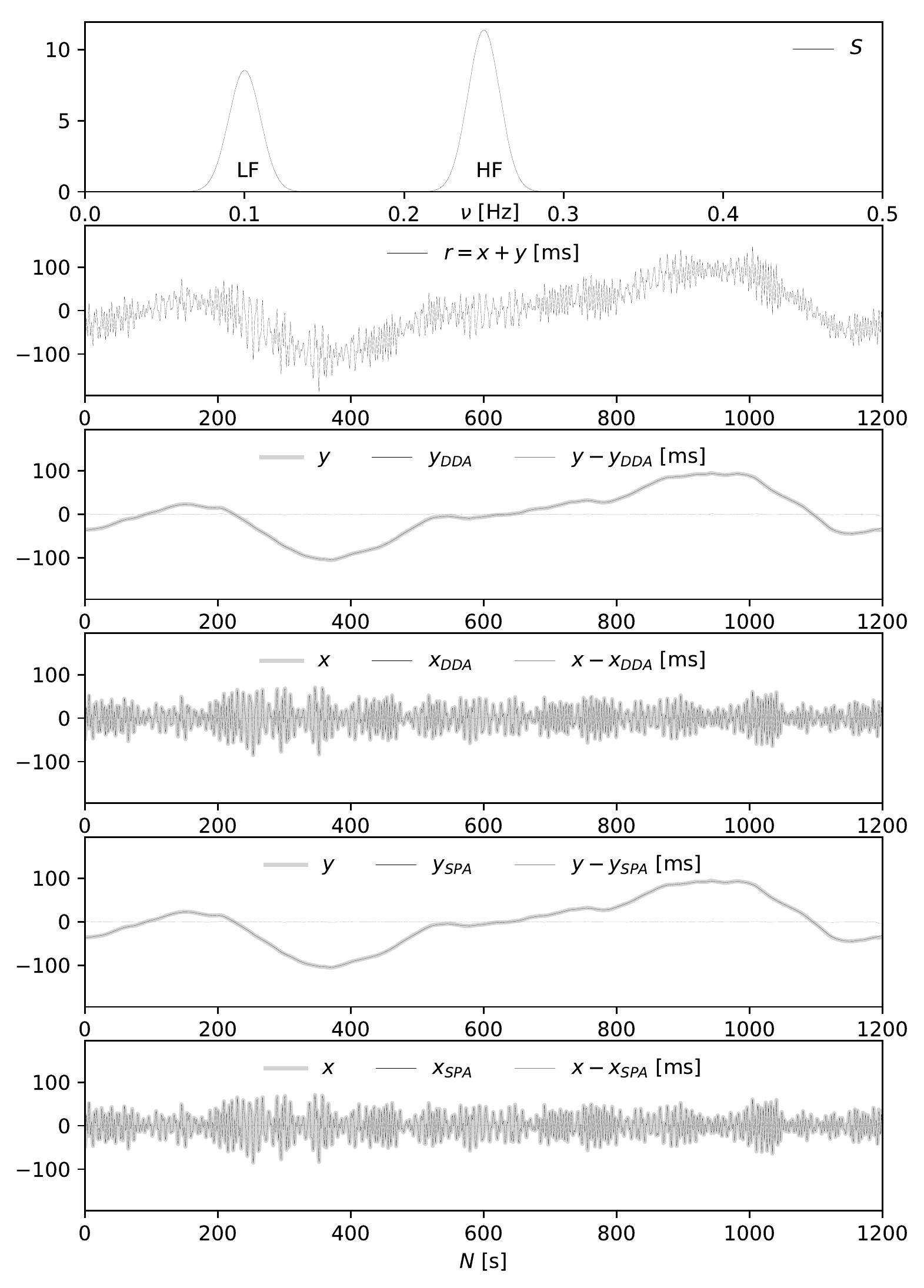}
\caption{Detrending example (see the text for discussion). 
}
\end{figure}

\section{Numerical examples}

The performance of the DDA method was estimated using 
synthetic RR time series generated with the model described in Ref. \cite{key-14}. 
In this model, a time series is generated with similar spectral characteristics 
to a real RR time series. The spectral characteristics of the RR time series,
including the low-frequency (LF) and high-frequency (HF) bands, are simulated 
using a bi-modal spectrum corresponding to the sum of two Gaussian functions: 
\begin{equation}
S(\nu) = \frac{P_{LF}}{\sqrt{2\pi}\sigma_{LF}} \exp\left( -\left[\frac{\nu-\nu_{LF}}{\sqrt{2}\sigma_{LF}}\right]^2\right) 
+ \frac{P_{HF}}{\sqrt{2\pi}\sigma_{HF}} \exp\left( -\left[\frac{\nu-\nu_{HF}}{\sqrt{2}\sigma_{HF}}\right]^2\right) 
\end{equation}
where $\nu_{LF}$, $\nu_{HF}$ and $\sigma_{LF}$, $\sigma_{HF}$ are means, and respectively the standard deviation, of each band. 
The power in the LF and HF bands is given by $P_{LF}$ and $P_{HF}$ respectively. 
The RR time series with the power spectrum $S(\nu)$ is generated
by taking the inverse Fourier transform of a sequence of
complex numbers with amplitudes $\sqrt{S(\nu)}$ and phases which
are randomly distributed in $[0,2\pi]$. 
By multiplying this time series with an appropriate constant and adding
the trend we obtain a realistic simulation of a RR time series. 
Typical values are 1000 ms for the average and 25 ms for standard deviation. 
Finally, the slowly varying trend is simulated by integrating a Gaussian random process.

An example of a spectrum $S(\nu)$ and the detrending of the corresponding time series $r(t)$ is given in Fig. 1. 
The power spectrum has the following parameters: 
$\sigma_{LF}=\sigma_{HF}=0.01$, $\nu_{LF} = 0.1$ Hz, $\nu_{HF}= 0.25$ Hz, $P_{LF}/P_{HF} = 0.75$. 
The $y_{DDA}$, $x_{DDA}$ are the trend and the stationary component recovered from $r=x+y$ using the 
DDA method, while $y_{SPA}$, $x_{SPA}$ are the trend and the stationary component 
recovered using the SPA method (the time series average was subtracted). 
The regularization parameter and the diffusion constant were set to $\mu = N$ and $\alpha = 0.25$. 
One can see that the DDA and SPA results are practically identical with the originals (similar results 
are obtained using the WSA method). 
We should note that the maximum diffusion constant value for which the finite difference solution 
is stable is $\alpha = 0.25$ (see Ref. \cite{key-15} for example). 

In Fig. 2 we give an estimation of the optimal regularization parameter $\mu$ using the root mean square error (RMSE) 
averaged over 1000 time series with the length $N=600$, which corresponds to 10 minutes of ECG recording. 
One can see that the optimal value is $\mu \simeq N$, for both SPA and DDA methods. We should note also that 
in this case the RMSE is smaller for the DDA method.

\begin{figure}[h!]
\centering \includegraphics[width=10.5cm]{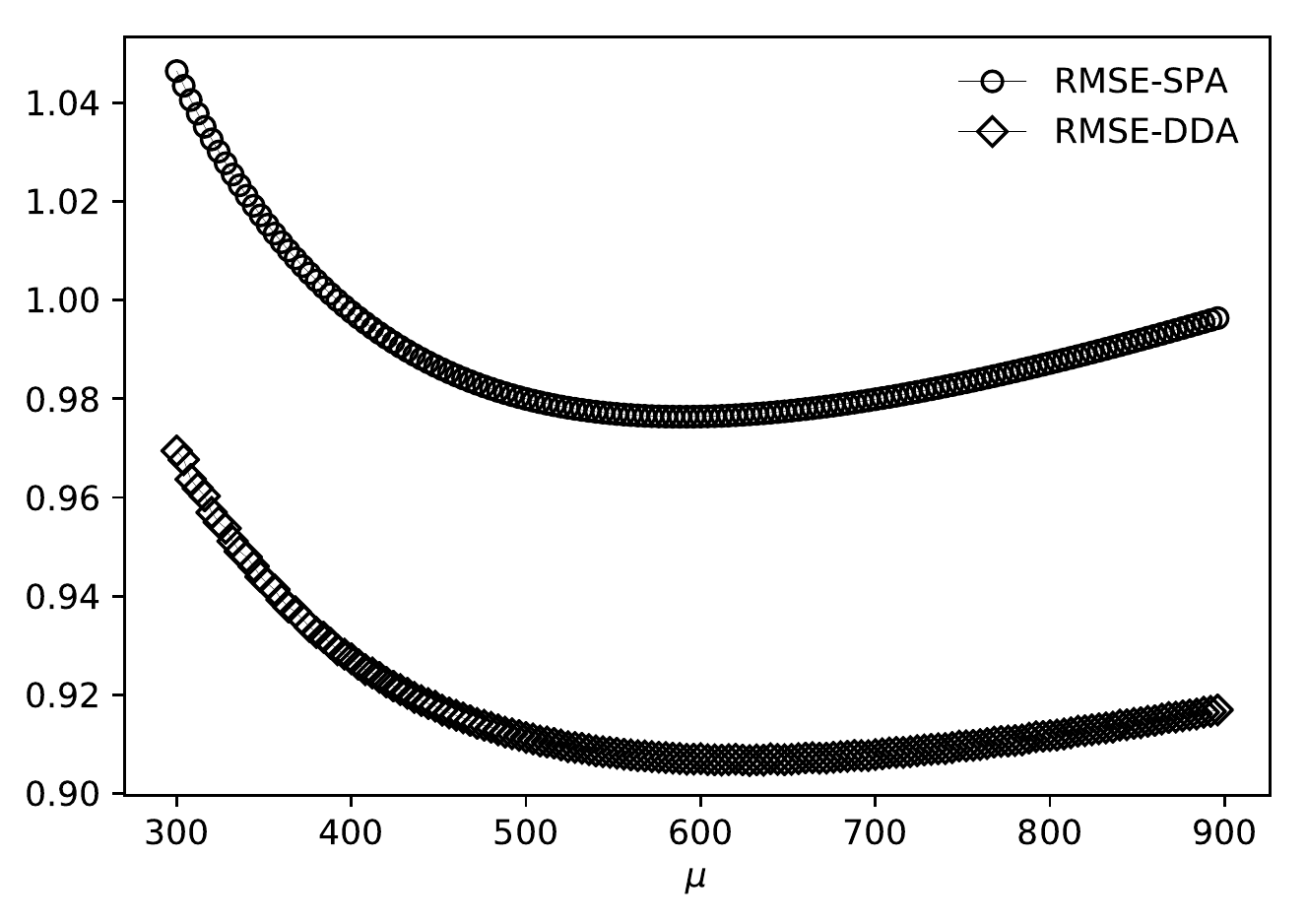}
\caption{Optimal regularization parameter $\mu$ as a function of $N$. 
}
\end{figure}

In order to evaluate the performance of the algorithms we used the ratio of the execution times $\rho=t_{SPA}/t_{DDA}$, $\rho=t_{WSA}/t_{DDA}$ and the RMSE 
as a function of the length $N$ of the time series (Fig. 3). The results are averaged over 1000 time series for each $N$ value. 
One can see that the ratio $\rho(N)$ increases with $N$, showing that the DDA method is about 250 times faster than the SPA method 
for time series with length $N=3600$, which corresponds to one hour ECG recordings. In the same time the WSA method is about twice faster than the DDA method, 
and this ratio does not depend on the length of the time series. 
The RMSE decreases with $N$, and the values for SPA, WSA and DDA are very close, with a slight advantage for DDA. 
The computations were performed on an Amazon EC2 instance with 64 GB RAM and 16 CPU cores at 3.4 GHz.

\begin{figure}[h!]
\centering \includegraphics[width=9cm]{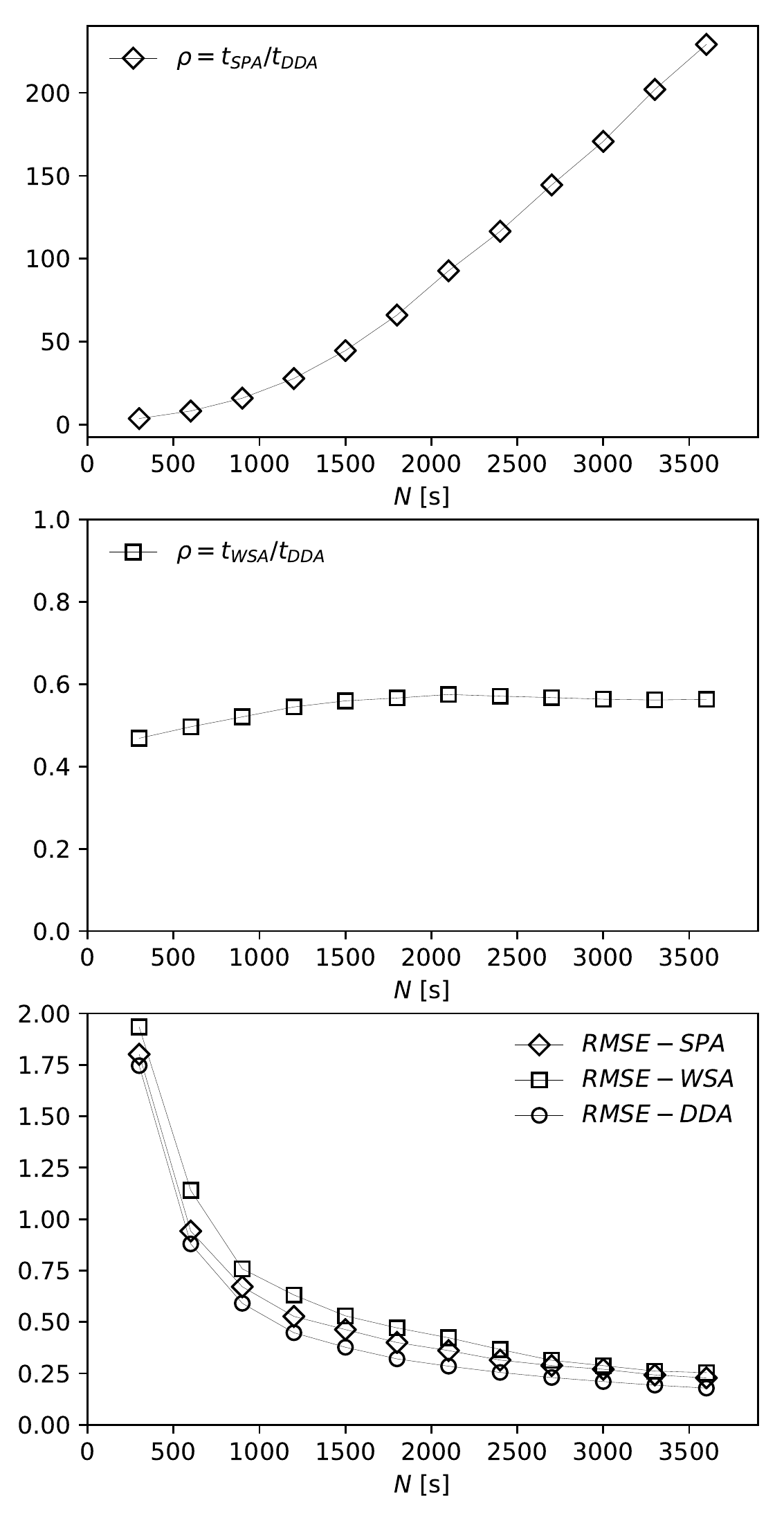}
\caption{The execution time ratio $\rho=t_{SPA}/t_{DDA}$, $\rho=t_{WSA}/t_{DDA}$ and the RMSE as a function of the RR timeseries length $N$. 
}
\end{figure}

The computation with the SPA method quickly becomes prohibitive for very long time series, however the DDA method has no problems 
scaling up, similarly to the WSA method. In Fig. 4 we show the execution time and the RMSE for time series with a length up to $N=86400$, corresponding to 
24 hours ECG recording (the results are averaged over 1000 time series). 
One can see that the computation takes only 0.5 seconds for 24 hours long time series using the DDA method.

\begin{figure}[h!]
\centering \includegraphics[width=9cm]{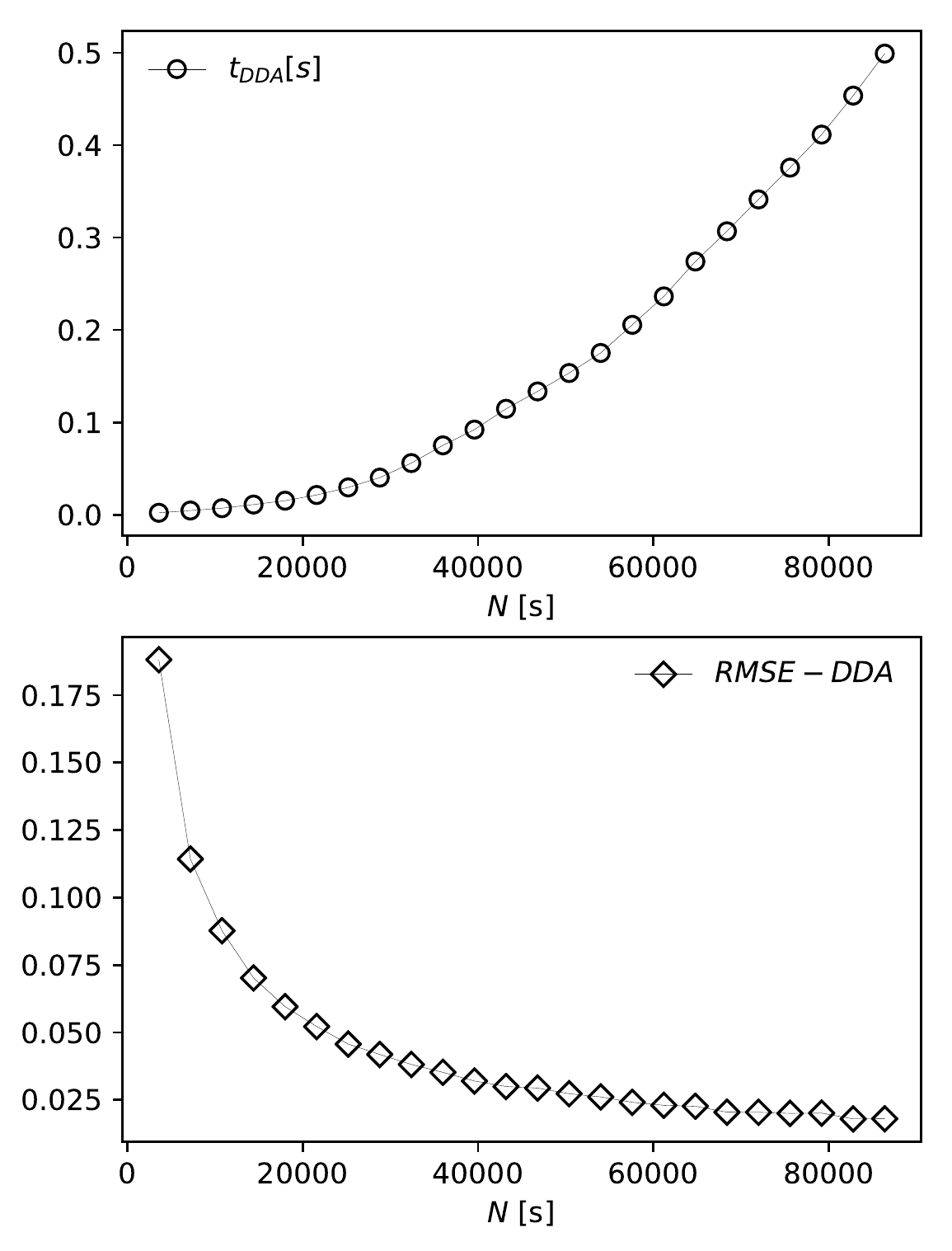}
\caption{The execution time $t_{DDA}$ and the RMSE as a function of the RR timeseries length for the DDA method up to 24 hours long time series. 
}
\end{figure}

\section*{Conclusion}

In this paper we presented a novel approach to detrending non-stationary RR time series. 
The approach is based on the finite difference iterative solution of the diffusion equation, with the initial condition 
given by the time series values. We have shown numerically that the diffusion method is equivalent to the SPA and WSA methods, which are widely used in HRV analysis. 
The speed of the proposed DDA method is comparable to the WSA method and it is several orders of magnitude faster than the SPA method, 
which makes it suitable for very long time series analysis. Besides the HRV examples discussed here, the described method can be used to accelerate any 
time series detrending, smoothing of fitting tasks. 

\section*{Appendix}

Here we provide the 'detrending.py', which is a Python3 implementation of the described algorithms (requires Numpy, Scipy),  
with the following functions:

\begin{itemize}
\item 
\begin{verbatim} 
def RR_timeseries(N, stdev) 
\end{verbatim}

Computes a synthetic RR time series with a given length N and standard deviation stdev 
(stdev=25 is recommended).

\item 
\begin{verbatim} 
def trend(N, k, a)
\end{verbatim}

Computes a synthetic trend with a given length N and amplitude a. 
The smoothness is controlled by the parameter k 
(k=2 is recommended).

\item 
\begin{verbatim} 
SPA_detrending(r, mu)
\end{verbatim}

Computes the SPA solution using the Cholesky method, given the timeseries r and the regularization parameter mu 
(mu=N is recommended).

\item 
\begin{verbatim} 
WSA_detrending(r, wavelet, mode, level)
\end{verbatim}

Computes the WSA solution, given the timeseries r, the thresholding mode (soft or hard) and the decomposition level. 
The implementation is based on the PyWavelets module.\cite{key-16}

\item 
\begin{verbatim} 
DDA_detrending(r, mu, alpha)
\end{verbatim}

Computes the DDA solution given the timeseries r, the regularization parameter mu, and the diffusion constant alpha 
(mu=N, alpha $\leq$ 0.25 are recommended).
 
\end{itemize}

\begin{small}
\begin{verbatim}
# detrending.py

import time
import pywt
import numpy as np
import scipy.linalg as sl

def RR_timeseries(N, stdev):
    N, df, ss = N//2, 1.0/N, 0.01*np.sqrt(2.0)
    alpha, fLF, fHF = 0.75, 0.1, 0.25
    s = np.zeros(N+1)
    for n in range(1, N):
        s[n] = alpha*np.exp(-np.power((n*df-fLF)/ss,2))
        s[n] += np.exp(-np.power((n*df-fHF)/ss,2))
    s = s*(N/np.sum(s))
    r = np.random.rand(N+1)*np.pi*2.0
    r = np.sqrt(s)*(np.cos(r) + np.sin(r)*1.0j)
    r = np.concatenate((r,np.conjugate(np.flip(r[1:N]))))
    r = np.real(np.fft.ifft(r))
    r = (stdev/np.std(r))*r
    return r, s

def trend(N, k, a):
    y = np.random.randn(N)
    for i in range(k):
        y = y - np.mean(y)
        for n in range(0,N-1):
            y[n+1] += y[n]
    y = (y-np.min(y))/(np.max(y)-np.min(y))
    return a*(y-np.mean(y))

def SPA_detrending(r, mu):
    N = len(r)
    D = np.zeros((N-2,N))
    for n in range(N-2):
        D[n,n], D[n,n+1], D[n,n+2] = 1.0, -2.0, 1.0
    D = mu*np.dot(D.T,D)
    for n in range(len(D)):
        D[n,n] += 1.0
    L = sl.cholesky(D,lower=True)
    Y = sl.solve_triangular(L,r,trans='N',lower=True)
    y = sl.solve_triangular(L,Y,trans='T',lower=True)
    return y, r - y

def WSA_detrending(r, wavelet, mode, level):
    coeff = pywt.wavedec( r, wavelet)
    tau = np.std(coeff[-level])*np.sqrt(2*np.log(len(r)))
    coeff[1:] = (pywt.threshold(i, value=tau, mode="soft") for i in coeff[1:])
    y = pywt.waverec( coeff, wavelet)
    return y, r-y

def DDA_detrending(r, mu, alpha):
    N, s, y, t = len(r), np.inf, np.copy(r), 0
    while True:
        f, g = y - r, y[0:N-2] + y[2:N] - 2*y[1:N-1]
        ss = np.dot(f,f) + mu*np.dot(g,g)
        if ss > s or t > N:
            break
        else:
            s, t = ss, t + 1
            y[1:N-1] = y[1:N-1] + alpha*g
            y[0] = y[0] + 2*alpha*(y[1] - y[0])
            y[N-1] = y[N-1] + 2*alpha*(y[N-2] - y[N-1])
    return t, y, r - y

if __name__ == '__main__':
    np.random.seed(54321)
    N = 1200
    x, s = RR_timeseries(N, 25)
    y = trend(len(x), 2, 200)
    r = x + y

    start_time = time.time()
    y_spa, x_spa = SPA_detrending(r, N)
    end_time = time.time()
    print('SPA execution time=', end_time - start_time, '[s]')
    print('SPA RMS=', np.linalg.norm(y-y_spa)/np.sqrt(N))

    start_time = time.time()
    y_wsa, x_wsa = WSA_detrending(r, "db32", "soft", 3)
    end_time = time.time()
    print('WSA execution time=', end_time - start_time, '[s]')
    print('WSA RMS=', np.linalg.norm(y-y_wsa)/np.sqrt(N))

    start_time = time.time()
    t, y_dd, x_dd = DDA_detrending(r, N, 0.25)
    end_time = time.time()
    print('DDA execution time=', end_time - start_time, '[s]')
    print('DDA RMS=', np.linalg.norm(y-y_dd)/np.sqrt(N))
\end{verbatim}
\end{small}

To run the program on a Linux computer use:

\begin{small}
\begin{verbatim}
python3 detrending.py
\end{verbatim}
\end{small}

The output will look like this:
\begin{small}
\begin{verbatim}
SPA execution time= 0.15083026885986328 [s]
SPA RMS= 0.8331844427187562
WSA execution time= 0.0009303092956542969 [s]
WSA RMS= 0.8856002765903443
DDA execution time= 0.002299070358276367 [s]
DDA RMS= 0.5105247820704635
\end{verbatim}
\end{small}

\end{document}